\documentclass[letter, oldversion]{aa}
\usepackage{graphicx}
\usepackage{txfonts}
\usepackage{aalongtable}
\usepackage{natbib}

\bibpunct{(}{)}{;}{a}{}{,}

\def\ltsima{$\; \buildrel < \over \sim \;$}
\def\simlt{\lower.5ex\hbox{\ltsima}} % < over ~
\def\gtsima{$\; \buildrel > \over \sim \;$}
\def\simgt{\lower.5ex\hbox{\gtsima}} % > over ~

\begin{document}

\idline{A\&A 1, 1--10 (2007)}{1} \doi{}

\title{The evolution of stars in the Taurus-Auriga T~association}
\author{C. Bertout\inst{1}
\and L. Siess\inst{2} \and S. Cabrit\inst{3}}

\institute{Institut d'Astrophysique, 98bis, Bd. Arago, 75014 Paris,
France \and Institut d'Astronomie et d'Astrophysique, Universit\'e
Libre de Bruxelles, CP 226, 1050 Brussels, Belgium  \and LERMA,
Observatoire de Paris, 61 Av. de l'Observatoire, 75014 Paris, France
}

\date{Received July 13, 2007   / Accepted }

\abstract
{In a recent study, individual parallaxes were determined for
many stars of the Taurus-Auriga T~association that are members
of the same moving group.
We use these new parallaxes to re-address the issue of the relationship
between classical T~Tauri stars (CTTSs) and weak-emission line
T~Tauri stars (WTTSs). With the available spectroscopic and photometric information for 72
individual stars or stellar systems among the Taurus-Auriga objects
with known parallaxes, we derived reliable photospheric
luminosities, mainly from the $I_c$ magnitude of these objects. We
then studied the mass and age distributions of the stellar sample,
using pre-main sequence evolutionary models to determine the basic
properties of the stellar sample. Statistical tests and Monte Carlo
simulations were then applied to studying the properties of the two
T~Tauri subclasses. We find that the probability of CTTS and WTTS samples being drawn from the same
parental age and mass distributions is low; CTTSs are, on average,
younger than WTTSs. They are also less massive, but this is due to
selection effects. The observed mass and age
distributions of both T~Tauri subclasses can be understood in the framework of a
simple disk evolution model, assuming that the CTTSs evolve into
WTTSs when their disks are fully accreted by the stars. According to
this empirical model, the average disk lifetime in Taurus-Auriga is
$4 \cdot 10^6 (M_*/M_\odot)^{0.75}$ yr.}
{} 

\maketitle

\section{Introduction} \label{Introduction}

In a recent work focusing on the kinematic properties of the
Taurus-Auriga T~association, \cite{2006A&A...460..499B} (Paper I hereafter) analyzed the
proper motions catalogued in \cite{2005A&A...438..769D}. They
identified a (minimum) moving group of 94 stars or stellar systems
sharing the same spatial velocity and derived kinematic parallaxes
for 67 of those objects. These co-moving stars define the
T-association as regards its kinematics.

In this Letter, we use these new parallaxes to re-address the
question of the relationship between the two subgroups of T~Tauri
stars (TTSs): (a) the CTTSs, which were detected primarily in the
course of H$\alpha$ surveys of dark clouds
and are actively accreting from their circumstellar disks, and (b)
the WTTSs that were identified both from H$\alpha$ and CaII line
emission surveys and X-ray surveys. They are magnetically active, young
stars that show no spectroscopic evidence
for accretion and are not usually associated with circumstellar
disks.

\begin{figure*}
\sidecaption
\includegraphics[width=12cm]{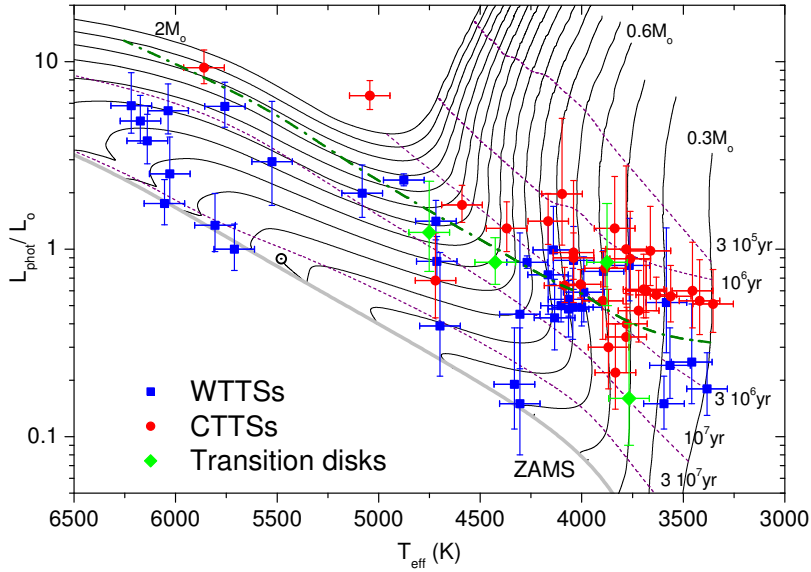}
\caption[]{HRD of the moving group stars. Red dots denote CTTSs,
blue squares mark WTTSs, and green diamonds indicate the stars with
transition disks (DI Tau, DM~Tau, GM~Aur, and Lk Ca 15). The error
bars indicate the $1\sigma$ uncertainties on $L_{phot}$ and
$T_{\!e\!f\!f}$. The solid black lines are evolutionary
tracks, computed with $Y=0.277$ and $Z=0.02$, for stars with masses
ranging from 0.3 to 2 $M_\odot$ with a mass increment of 0.1
$M_\odot$. The current position of the Sun as computed with these
parameters is shown as a solar symbol. Note that the parameters chosen
for the computation of pre-main sequence evolution tracks are not solar,
so that the properties of the computed 1 $M_\odot$ model do not exactly correspond to those of the actual Sun.
The dashed lines are isochrones for the ages indicated in the figure,
while the heavy dark-green dash-dotted line shows the disk lifetime
$\tau_d$ as a function of stellar mass as defined by Eq.~\ref{Eq1}.} \label{HRD}
\end{figure*}

The exact relationship between CTTSs and WTTSs remains elusive. Both
species share the same region of the Hertzsprung-Russell diagram
(HRD), and the most obvious difference
between them is their somewhat different location with respect to
the molecular clouds with which they are associated. The new
parallaxes derived in Paper I have confirmed the
previous finding \citep [e.g., ][and references
therein]{1996ApJ...468..306F} that the distribution in space of
optically detected TTSs and X-ray selected WTTSs is different. CTTSs
are usually found in the inner regions of the molecular clouds,
while WTTSs tend to be found on the outskirts of starforming
regions. However, optical WTTSs were originally searched for only in
the immediate vicinity of the molecular clouds and were therefore found in
roughly the same regions as CTTSs. In contrast, X-ray selected WTTSs
were mainly discovered in the course of large-scale X-ray surveys,
so some of them were found far away from the central parts of
starforming regions. More recent searches for additional optical WTTSs
did not lead to the discovery of many previously unnoticed WTTSs in the
Taurus-Auriga area. While successful at discovering
new CTTSs in Taurus, the \emph{Spitzer} observations reported by
\cite{2006ApJ...647.1180L} show that the previous census
\citep{1993AJ....105.1927G} of Taurus objects down to $V \approx 14$
-- which is also the approximate limiting
magnitude of the \cite{2005A&A...438..769D} sample -- was 80\%
complete.

To explain the existence of both CTTSs and WTTSs, one usually
postulates that newborn stars display a wide range of disk masses
and that their accretion or dispersal requires a correspondingly
wide range of timescales, which justifies the stars intermingling in
the HRD \citep[e.g.,][]{1995ApJ...450..824S}. In other terms,
it is only the evolutionary status of their
circumstellar disks that distinguishes the subgroups. However, this
explanation does not rest on firm observational support yet.
We thus re-address here the issue of the respective evolutionary
status of CTTSs and WTTSs. We show that there are in
fact significant differences in their ages and that their mass and
age distributions can be understood within a simple evolutionary
framework.

\section{Photospheric luminosities of moving-group stars}  \label{BasicData}

In Paper I, we derived parallaxes for 30 CTTSs or
CTTS systems, 36 WTTSs or WTTS systems, and 1 Herbig Be star. The respective
numbers of CTTSs and WTTSs given here correct those given in Sect.
6.3 of Paper I, where some objects were
misidentified. The conclusions of Paper I are not affected by this error.
We mainly used the photometric and spectroscopic data given by
\cite{1995ApJS..101..117K} in their thorough investigation of the
Taurus-Auriga population.  The visual extinction values were also
taken mainly from that work. We supplemented this material with
newer data for WTTSs that were discovered after 1995 and
for a number of system components that were resolved by recent high
angular-resolution observations. To avoid a selection bias, we did
not include the components of stellar systems with spectral types
later than M4 or derived luminosities lower than $\approx 0.15
L_\odot$; these values correspond to the coolest and faintest,
apparently single objects found in the original sample. For reasons
discussed by \cite{2005ApJ...635..422C}, we also excluded those CTTS
components for which only $J$ or $K$ flux measurements are
available. Our final sample, including the resolved components of
multiple systems that fulfill the above criteria, comprises 33 CTTSs
and 38 WTTSs.

\renewcommand{\arraystretch}{1.2}
\begin{table}
\caption{Probabilities that the distributions of stellar properties
of CTTSs and WTTSs are drawn from the same parent
distribution.}\label{KSResults1}
\begin{center}
\small{
\begin{tabular}{lc}
\hline \hline
%\multicolumn{2}{c} {Observed properties} \\
%\hline
%
%$V$-mag & $4.0 \cdot 10^{-4}$   \\
%De-reddened $V$-mag &  0.17    \\
%$T_{eff}$ & $4.7 \cdot 10^{-4}$ \\
%
%\hline
%\multicolumn{2}{c} {Derived $L_*$ for data and control samples} \\
%\hline
%
%$L_{*}$ & $0.50 \pm 0.23$  \\
%$L_{*}$ (CS1) & $0.92$ \\
%$L_{*}$ (CS2) & $0.85$  \\
%
%\hline
\multicolumn{2}{c} {Derived masses for data and control samples}  \\
\hline
Mass &  $(3.9 \pm 3.4)\cdot 10^{-4}$    \\
Mass (CS1) &  $(7.8 \pm 5.1)\cdot 10^{-4}$   \\
Mass (CS2) &  $(2.4 \pm 2.1)\cdot 10^{-4}$ \\
\hline
\multicolumn{2}{c} {Derived ages for data and control samples}  \\
\hline
Age & $(1.4 \pm 1.2)\cdot 10^{-4}$  \\
Age (CS1) & $(3.7 \pm 2.0)\cdot 10^{-3}$  \\
Age (CS2) & $(1.4 \pm 0.8)\cdot 10^{-2}$  \\
\hline
\end{tabular}
}
\end{center}
\end{table}
\renewcommand{\arraystretch}{1.0}

The stellar luminosities of all CTTSs were computed from their $I_C$
flux, because the contribution of excess emission is smallest in
that filter \citep[see][]{2005ApJ...635..422C} and from the
appropriate bolometric correction, as tabulated for various spectral
types in \cite{1995ApJS..101..117K}. For the WTTSs, we derived the
luminosities either from the $I_C$ flux when available or from the
$V$ flux. Since WTTSs have negligible excesses in the optical range,
both procedures should be equivalent. Stars for which both
measurements were available indeed give the same results to
within a few percent. In a few cases of WTTSs lacking
visual-extinction determination, we computed an approximate value by
requiring that the $V$ and the $I_C$ (when available) or 2MASS $J$
flux measurements give the same stellar luminosity.
%Error bars on the luminosities were computed by considering the various
%uncertainties on the quantities entering their determination. We found that
The uncertainties on the new parallaxes dominate the
luminosity error budget so strongly that other sources of error can be
neglected in a first approximation.

Finally, we built two control samples for the same objects by (a)
computing the luminosities as explained above but assuming that the
stars were all located at the post-Hipparcos average distance of
$139^{+12}_{-10}$ pc derived for Taurus-Auriga by
\cite{1999A&A...352..574B} (called sample CS1 hereafter) and (b)
computing the luminosities from the 2MASS $J$ flux and assuming the
same distance of $139^{+12}_{-10}$ pc for all stars (sample CS2
hereafter). The CS1 sample allows us to gauge the effect of the new
parallaxes, while CS2 allows for a direct comparison of our result
with those of \cite{1995ApJS..101..117K}.

\section{HRD of the moving group}\label{HRDSection}

We used the grid of pre-main sequence evolutionary tracks computed
by \cite{2000A&A...358..593S} to derive the masses and age of our
stars along with error bars due to the uncertainty on the
luminosities and effective temperatures (we assumed a $\pm 100$ K
uncertainty for all spectral types). Online Table~\ref{Properties} gives the basic
stellar properties of our sample and their uncertainties.

\renewcommand{\arraystretch}{1.2}
\begin{table}
\begin{center}
\caption{Average and median masses and ages for TTS
samples.}\label{AverMT}
\small{
\begin{tabular}{lcccc}
\hline\hline &  $\overline{M} \pm \sigma_{\overline{M}}$ & Med. $M$
&  $\overline{\log t} \pm \sigma_{\overline{\log t}}$ & Med.$\log t$  \\
 & M$_\odot$ & M$_\odot$ & $t$ in yr & $t$ in yr \\
\hline
CTTSs  &   $0.72  \pm  0.44$   &  $0.57$ &  $6.37 \pm 0.37$ & $6.28$ \\
WTTSs  &   $0.97  \pm  0.40$   &  $0.80$ &  $6.82 \pm 0.44$ & $6.74$ \\
%OWTTSs  &  $0.86  \pm  0.35$   &  $0.78$ &  $6.59 \pm 0.32$ & $6.59$ \\
%XWTTSs  &  $1.12  \pm  0.39$   &  $1.20$ &  $7.05 \pm 0.50$ & $7.11$ \\
%
\hline
\end{tabular}
}
\end{center}
\end{table}
\renewcommand{\arraystretch}{1.0}

It is apparent in the HRD of the moving group plotted in
Fig.~\ref{HRD} that the CTTSs appear on average younger and less
massive than WTTSs. This is confirmed by the Kolmogorov-Smirnov (K-S) statistics for the
masses and ages of the data sample shown in Table~\ref{KSResults1}.
As discussed by \cite{2001ASPC..243..581S}, pre-main
sequence evolutionary tracks computed by different groups and using
different input physics yield different masses and ages for a given
pre-main sequence position in the HRD. The uncertainties on $T_{\!e
\!f \!f}$ and $L_{phot}$ adopted in this study are large enough to
offset the age and mass uncertainties that arise
from the physical approximations made in our evolutionary code, but
do not account for possible systematic effects, such as those
introduced by the different shapes of evolutionary tracks
obtained by different investigators. To assess the robustness of the
mass and age determinations from our models, we thus re-computed
the masses and ages of both TTS subgroups by using the
\cite{1997MmSAI..68..807D} tracks and isochrones, and
found that the resulting K-S probabilities for the CTTSs and WTTSs
to be drawn from the same mass and age parental distributions are
respectively $8.3 \cdot 10^{-3}$ and $6.6 \cdot 10^{-5}$. This
result increases our confidence that the differences between CTTSs
and WTTSs reported here are not an artefact caused by our
evolutionary tracks. We nevertheless caution that the mass and age values given in
Table~\ref{AverMT} and the average disk lifetime derived in
Sect.~\ref{Evolution} depend on the evolutionary models used in this
investigation.

\begin{figure}
\resizebox{\hsize}{!}{\includegraphics[angle=0]{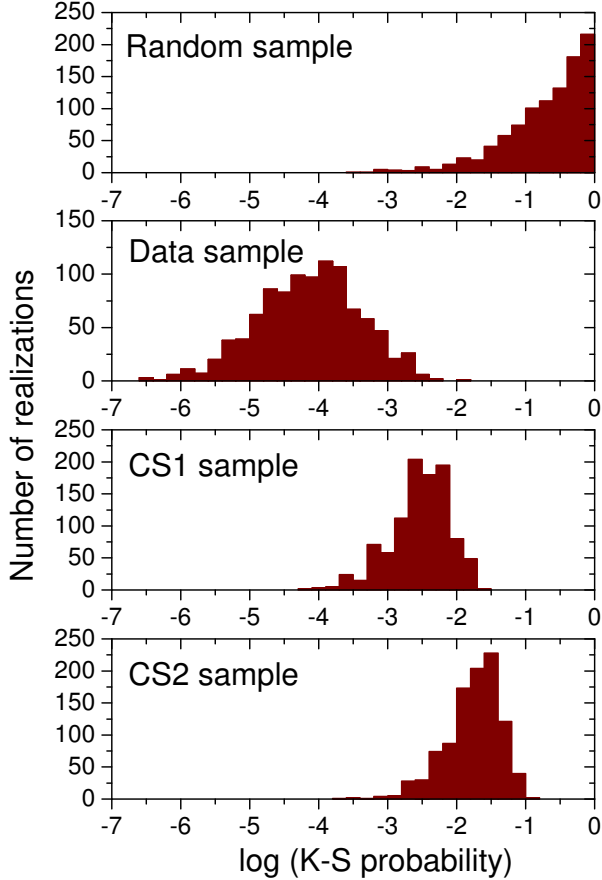}}
\caption[]{K-S probability histograms from Monte Carlo simulations
for the age distributions of CTTSs and WTTSs (see text for
details).} \label{KSAges}
\end{figure}

The difference in average masses between the two TTS subgroups is
due to an observational selection effect, since the luminosity limit
of $\approx 0.15 L_\odot$ for our sample precludes any star with a
mass lower than $\approx 0.4 M_\odot$ being older than $\approx 5
\cdot 10^6$ yr. Because this is an age range for which we expect to
have more WTTSs than CTTSs, the mass distribution of WTTSs is skewed
toward high masses. This is confirmed by a K-S test showing that the
probability is 0.27 that CTTSs and WTTSs more massive than
$0.7M_\odot$ are drawn from the same distribution.

The age distributions for CTTSs and WTTSs do not suffer from such a
bias, since the missing low-mass WTTSs would only reinforce the age
difference between CTTSs and WTTSs suggested by the K-S statistics
of Table~\ref{KSResults1}. Note that the probability that
the age distributions are drawn from the same parental distribution
is $\sim 10^{-3}$ for the CS1 control sample and $\approx 10^{-2}$
for the CS2 sample.  This second value is less significant by two
orders of magnitude than the K-S probability found for our data
sample using the new parallaxes ($\approx 10^{-4}$), but it agrees
with what \cite{1995ApJS..101..117K} found in their own study.

We performed Monte-Carlo (MC) simulations to assess the significance of
this finding. We first constructed 1000 realizations of our dataset
by varying the effective temperatures and computed luminosities
within their uncertainties, which we assumed to be normally
distributed. We then computed the CTTS and WTTS age distributions
for each realization and computed their K-S statistics. The result
is a histogram of K-S probabilities for the various realizations of
the data set. We repeated this procedure for our two control
samples. As a final test, we constructed one more control
sample by randomly drawing the effective temperature and stellar
luminosity of individual stars within their observed (non-normal)
distributions and computing masses and ages of our sample using
these values. The histogram of K-S probabilities for this ``random''
dataset thus allows us to check the validity of the K-S test when
used with our non-normal dataset.
\renewcommand{\arraystretch}{1.2}
\begin{table*}
\begin{center}
\caption{Number of MC realizations with K-S
probabilities higher than 0.9995
(see text) and average values of the
parameters (with their standard deviations) for 1000
realizations of each investigated sample.}\label{KSDistrib}
\small{
\begin{tabular}{lccccc}
\hline\hline   & $N(p_{K-S} > 0.9995)$ & $\overline{\log
(\alpha/\gamma)_{CTTSs}} \pm \sigma$ &  $\overline{\beta_{CTTSs}}
\pm \sigma$ & $\overline{\log
(\alpha/\gamma)_{WTTSs}} \pm \sigma$ &  $\overline{\beta_{WTTSs}} \pm \sigma$  \\
\hline
Data  &   97.6\% & $6.63  \pm  0.05$   &  $2.85 \pm 0.21$ &  $6.58  \pm  0.08$   &  $2.87 \pm 0.21$ \\
CS1   &   59.2\% & $6.62  \pm  0.03$   &  $2.90 \pm 0.15$ &  $6.58  \pm  0.05$   &  $2.76 \pm 0.22$ \\
CS2   &   5.2\% & $6.33  \pm  0.06$   &  $2.21 \pm 0.17$ &  $6.26  \pm  0.04$   &  $2.03 \pm 0.18$ \\
Random&   8.0\% & $6.78  \pm  0.13$   &  $3.89 \pm 0.61$ &  $6.76  \pm  0.12$   &  $3.89 \pm 0.54$ \\
\hline
\end{tabular}
}
\end{center}
\end{table*}
\renewcommand{\arraystretch}{1.0}
Figure~\ref{KSAges}  displays the 4 resulting histograms of K-S
probabilities on the same logarithmic scale. The data and random
sample histograms have very little overlap, which confirms the
significance of the age difference between CTTSs and WTTSs. Age
differences are also apparent in the CS1 and CS2 samples, but with
lower significance. Clearly, the individual parallaxes of
moving-group stars used in this investigation allow for a better
differentiation of the two subgroups, even though the large parallax
uncertainties act to broaden the probability histogram when compared
to the control samples.

\section{The evolution from CTTSs to WTTSs} \label{Evolution}

With this first indication that a distinction between CTTSs and
WTTSs can be seen in the HRD, we were encouraged to look for a
simple empirical model that would explain the evolution from CTTS to
WTTS based on disk evolution. We thus assumed that the mass of a
T~Tauri disk in units of $M_\odot$ is given by $M_d = \alpha
(M_*/M_\odot)^\beta$ and that the accretion rate is given by $\dot
M_d = \gamma (M_*/M_\odot)^{2.1}$ $M_\odot$/yr, in agreement with
current observations over a wide mass range
\citep{2003ApJ...592..266M, 2005MmSAI..76..343N}. From there, we
derived the disk lifetime $\tau_d$ (in yr)
\begin{equation} \label{Eq1}
\log \tau_d = log (\alpha/\gamma) + (\beta - 2.1) \log
(M_*/M_\odot).
\end{equation}
For a given combination of $\alpha/\gamma$ and $\beta$, we then
compared the age $t_*$ of each star -- as derived from its
evolutionary track -- to the disk age $\tau_d(M_*)$. We then assumed
that the star was a CTTS whenever $t_* \leq \tau_d(M_*)$ or a WTTS
whenever $t_*> \tau_d(M_*)$. Once we assigned a type to all stars in
the sample, we compared the resulting distributions of model CTTS
and WTTS masses to the observed mass distributions. Using the K-S
statistics to indicate goodness-of-fit, we looked for the best
possible match between the mass distributions while varying the
parameters in the range $5 \leq \log (\alpha/\gamma) \leq 8$ and $-4
\leq \beta \leq 4$. Much to our surprise -- since our model is so
simple -- we found excellent statistical agreement between the
observed and modeled mass distributions for both CTTSs and WTTSs
\emph{in a single and very well-defined range of parameters} around
$\log (\alpha/\gamma) \approx 6.6$ and $\beta \approx 2.9$. The
resulting $\tau_d$ locus in the HRD is shown in Fig.~\ref{HRD}.

To examine this in more detail, we again resorted to MC simulations,
scanning the $(\alpha/\gamma, \beta)$ parameter
space for each of the 1000 realizations of our data sample. The
results are summarized in Table~\ref{KSDistrib} with the results
found for the CS1, CS2, and random samples (see above). Again, the
derived $\alpha/\gamma$ and $\beta$ are confined to a single and
well-defined range of values. We find that 8\% of the random sample
have a K-S probability 0.9995 or higher that the
modeled and observed CTTS and WTTS mass distributions are drawn from the same
respective parental distributions. In contrast, this probability is higher than
0.9995 in \emph{97.6\%} of all realizations of our actual data set,
and the computations yield very similar parameter values for both
CTTSs and WTTSs. Simulations using the CS1 sample gave a lower
percentage of high-probability results, but the derived parameter
values are very close to those found with the original dataset. Our
results thus remain valid if we use the Taurus
average parallax value instead of the individual parallaxes.
The situation is much less favorable when
using the CS2 sample, since the number of high-probability
realizations drops down to a value similar to the one for the random
sample. This confirms that using the flux in $J$ as a proxy
for the photospheric luminosity of CTTSs blurs the differences
between CTTS and WTTS properties.

We conclude that the observed distributions of ages and masses
\emph{in the Taurus-Auriga moving group} can be explained by
assuming that a CTTS evolves into a WTTS when the disk is fully
accreted by the star. Such an evolution has been hypothesized for a
long time, but it is the first time that observational evidence
unambiguously supports this scenario. The average disk lifetime in
Taurus-Auriga is found to be $4 \cdot 10^6 (M_*/M_\odot)^{0.75}$ yr
in the framework of our heuristic. If we assume an average mass
accretion rate of $10^{-8} M_\odot$/yr for a typical $1M_\odot$
CTTS, the disk mass is given in the same framework by $M_d = 0.04
(M_*/M_\odot)^{2.85}$.

The results derived above are only valid in a statistical sense.
As seen in Fig.~\ref{HRD}, some CTTSs
are older than the lifetime of their disk, while some WTTSs
are younger; these stars are thus misidentified by our model.
The misidentified CTTSs are the oldest members of the
subgroup: CW~Tau, GO~Tau, DM~Tau, VY~Tau, IP~Tau, LkCa~15, and
GM~Aur. Only 2 CTTSs (GO~Tau and CW~Tau) out of 33 are older by more
than $1\sigma$ than their expected disk lifetime $\tau_d$. Mass
accretion for DM Tau, GM~Aur, and IP~Tau range from $8 \cdot
10^{-10}$ to $9 \cdot 10^{-9} M_\odot$/yr
\citep{1993AJ....106.2024V,1998ApJ...492..323G}, thus indicating
rather mild accretion activity. Furthermore, DM~Tau, GM~Aur, and
LkCa~15 are believed to be transition objects that are currently
clearing their disks \citep{2005ApJ...630L.185C,
2006A&A...460L..43P}. Properties of these objects are thus
compatible with those of CTTSs approaching the end of their
disk-accretion phase. The exception that our model does not account
for is CW~Tau, which harbors a jet and has a mass accretion rate of
$6 \cdot 10^{-8}M_\odot$/yr \citep{2005ApJ...626..498M}. The
misidentified WTTSs are also the youngest of the sample: LkCa~14,
V928~Tau, DI~Tau, IW~Tau, V827~Tau, and RXJ0432.8+1735. Again, most
of these objects have ages within $1\sigma$ of their respective
$\tau_d$. While most of them are rather unremarkable WTTSs,
\cite{1997ApJ...489L.173M} have proposed that DI~Tau is a
transitional object surrounded by a fast-evolving disk.

As a final remark, we note that the relationship between disk mass
and stellar mass derived above leads to very low masses for disks
surrounding brown dwarfs and will thus be falsified by the first
observation of a massive brown dwarf disk  \emph{unless the masses
of CTTS disks have been largely underestimated}, which remains a
possibility if substantial grain growth takes place in these disks
\citep[cf.][]{2007prpl.conf..767N}. As mentioned by
\cite{2006ApJ...639L..83A} and  \cite{2006ApJ...648..484H}, a steep
relationship between disk and stellar masses would also help us
understand the dependence of the mass accretion rate on the stellar
mass.

%\begin{acknowledgements}
%This research made use of the Centre de Donn\'ees de Strasbourg
%facilities and of the NASA/ IPAC Infrared Science Archive.
%\end{acknowledgements}

\bibliographystyle{aa}
\bibliography{8276}
%\Online
\renewcommand{\arraystretch}{1.2}
\small{
\renewcommand{\footnoterule}{}  % to avoid a line before footnotes
\begin{longtable}{lccccccccc}
\caption{\label{Properties} Observational and derived properties of moving-group stars used in this study.} \\
\hline\hline Star    &   Type\footnotemark[1]    &   Sp. Type    &
$T_{\!e\!f\!f}$   &
$A_V$   &   $L_{star}/L_\odot$     & Filter\footnotemark[2] &   $M_{star}/M_\odot$&   $\log t \pm \sigma_{\log t}$ (t in yr)  & Ref.\footnotemark[3]       \\
\hline
\endfirsthead
\caption{continued.}\\
\hline\hline Star    &   Type\footnotemark[1]    &   Sp. Type    &   $T_{\!e\!f\!f}$   &   $A_V$   &   $L_{star}/L_\odot$     & Filter\footnotemark[2] &   $M_{star}/M_\odot$          &   $\log t \pm \sigma_{\log t}$ (t in yr)   & Ref.\footnotemark[1]       \\
\hline
\endhead
\hline
\endfoot
\noalign{\smallskip}\hline \noalign{\medskip}
\multicolumn{10}{@{}l@{}}{$^1$Meaning of symbols: c = CTTS, o = optically selected WTTS, x = X-ray selected WTTS.}\\
\multicolumn{10}{@{}l@{}}{$^2$Photometric filter used for determining the photospheric luminosity. I stands for the Cousins I filter, J for the 2MASS J filter.} \\
\multicolumn{10}{@{}l@{}}{\hspace{0.8mm} When 2 filters are given, $A_V$ is computed by assuming that the photospheric luminosities using both filters are equal.}\\
\multicolumn{10}{@{}l@{}}{$^3$References for observational properties: (1) \cite{1995ApJS..101..117K}; (2) \cite{1988AJ.....96..297W}; (3) \cite{2000A&A...359..181W};}\\
\multicolumn{10}{@{}l@{}}{\hspace{0.8mm}  (4) \cite{1995AJ....110..776W}; (5) \cite{2001A&A...376..982W}; (6) \cite{1999ApJ...520..811W}; (7) \cite{1993A&A...278..129L}; (8) \cite{1997ApJ...490..353G}.}\\
\endlastfoot
NTTS 035120+3154SW  &   x   &   G0  &   6030    &   0.87    &   2.52    $^{+    1.44    }_{ -0.77   }$  &   I   &   1.27    $\pm$   0.08    &   7.78    $\pm$   0.91    &   1   \\
NTTS 035120+3154NE  &   x   &   G5  &   5770    &   0.97    &   1.34    $^{+    0.64    }_{ -0.37   }$  &   I   &   1.14    $\pm$   0.05    &   7.43    $\pm$   0.08    &   1   \\
NTTS 040047+2603E   &   x   &   M2  &   3580    &   0.49    &   0.25    $^{+    0.26    }_{ -0.1    }$  &   I   &   0.4 $\pm$   0.04    &   6.4 $\pm$   0.19    &   1   \\
GSC 01262-00421 &   x   &   F8  &   6200    &   0:    &   4.82    $^{+    1.79    }_{ -1.15   }$  &   V   &   1.44    $\pm$   0.05    &   7.14    $\pm$   0.06    &    2   \\
RX J0405.7+2248 &   x   &   G0  &   6030    &   0.02    &   5.47    $^{+    2.15    }_{ -1.35   }$  &   V,I &   1.52    $\pm$   0.1 &   7.07    $\pm$   0.08    &   3   \\
RX J0406.7+2018 &   x   &   F8  &   6200    &   0.05    &   3.78    $^{+    1.46    }_{ -0.93   }$  &   V,I &   1.39    $\pm$   0.06    &   7.22    $\pm$   0.06    &   3   \\
%V773 Tau ABCD   &   m   &   K3  &   4730    &   1.32    &   5.21    $^{+    0.41    }_{ -0.37   }$  &   I   &   1.95    $\pm$   0.13    &   6.12    $\pm$   0.09    &   1   \\
V773 Tau A  &   o   &   K2  &   4900    &   1.32    &   2.34    $^{+    0.18    }_{ -0.16   }$  &   I   &   1.66    $\pm$   0.05    &   6.69    $\pm$   0.07    &    4   \\
V773 Tau B  &   o   &   K5  &   4350    &   1.32    &   0.85    $^{+    0.07    }_{ -0.06   }$  &   I   &   1.13    $\pm$   0.08    &   6.71    $\pm$   0.1 &    4   \\
%V773 Tau C  &   c   &   M0  &   3850    &   1.32    &   0.53    $^{+    0.04    }_{ -0.04   }$  &   J   &   0.57    $\pm$   0.06    &   6.36    $\pm$   0.1 &       \\
CW Tau  &   c   &   K3  &   4730    &   2.19    &   0.68    $^{+    0.54    }_{ -0.25   }$  &   I   &   1.11    $\pm$   0.15    &   7.21    $\pm$   0.22    &   1   \\
FP Tau  &   c   &   M4  &   3370    &   0.24    &   0.51    $^{+    0.27    }_{ -0.15   }$  &   I   &   0.31    $\pm$   0.03    &   6.04    $\pm$   0.14    &   1   \\
CX Tau  &   c   &   M2.5    &   3580    &   0.83    &   0.56    $^{+    0.26    }_{ -0.15   }$  &   I   &   0.4 $\pm$   0.04    &   6.12    $\pm$   0.09    &   1   \\
LkCa 4  &   o   &   K7  &   4060    &   0.69    &   0.73    $^{+    0.65    }_{ -0.28   }$  &   I   &   0.77    $\pm$   0.09    &   6.43    $\pm$   0.25    &   1   \\
CY Tau  &   c   &   M1  &   3720    &   0.1 &   0.4 $^{+    0.09    }_{ -0.07   }$  &   I   &   0.48    $\pm$   0.05    &   6.37    $\pm$   0.11    &   1   \\
LkCa 5  &   o   &   M2  &   3580    &   0.1 &   0.24    $^{+    0.14    }_{ -0.08   }$  &   I   &   0.39    $\pm$   0.04    &   6.5 $\pm$   0.17    &   1   \\
NTTS 041529+1652    &   x   &   K5  &   4350    &   0   &   0.15    $^{+    0.23    }_{ -0.07   }$  &   I   &   0.85    $\pm$   0.19    &   7.5 $\pm$   0.37    &   1   \\
V410 Tau ABC    &   o   &   K4  &   4730    &   0.03    &   1.41    $^{+    0.41    }_{ -0.29   }$  &   I   &   1.45    $\pm$   0.09    &   6.8 $\pm$   0.12    &   1   \\
DD Tau AB   &   c   &   M1  &   3720    &   1.61    &   0.61    $^{+    0.33    }_{ -0.18   }$  &   I   &   0.48    $\pm$   0.05    &   6.16    $\pm$   0.13    &   1   \\
CZ Tau AB   &   c   &   M1.5    &   3650    &   1.32    &   0.6 $^{+    0.28    }_{ -0.17   }$  &   I   &   0.43    $\pm$   0.04    &   6.13    $\pm$   0.11    &   1   \\
Hubble 4    &   o   &   K7  &   4060    &   0.76    &   0.48    $^{+    0.24    }_{ -0.14   }$  &   V   &   0.79    $\pm$   0.08    &   6.72    $\pm$   0.21    &    1   \\
NTTS 041559+1716    &   x   &   K7  &   4060    &   0   &   0.45    $^{+    0.77    }_{ -0.22   }$  &   I   &   0.79    $\pm$   0.08    &   6.74    $\pm$   0.4 &   1   \\
BP Tau  &   c   &   K7  &   4060    &   0.49    &   0.65    $^{+    0.13    }_{ -0.1    }$  &   I   &   0.78    $\pm$   0.08    &   6.51    $\pm$   0.12    &   1   \\
V819 Tau AB &   o   &   K7  &   4060    &   1.35    &   0.43    $^{+    0.26    }_{ -0.14   }$  &   I   &   0.8 $\pm$   0.08    &   6.79    $\pm$   0.24    &   1   \\
%LkCa7 AB    &   m   &   K7  &   4060    &   0.59    &   0.67    $^{+    0.36    }_{ -0.2    }$  &   I   &   0.78    $\pm$   0.09    &   6.49    $\pm$   0.2 &   1   \\
LkCa7 A &   o   &   K7  &   4060    &   0.59    &   0.59    $^{+    0.32    }_{ -0.18   }$  &   J   &   0.78    $\pm$   0.08    &   6.57    $\pm$   0.21    &   5    \\
LkCa7 B &   o   &   M3.5    &   3420    &   0.59    &   0.18    $^{+    0.1 }_{ -0.05   }$  &   J   &   0.31    $\pm$   0.03    &   6.5 $\pm$   0.13    &    5   \\
RY Tau  &   c   &   K1  &   5080    &   1.84    &   6.59    $^{+    1.35    }_{ -1.04   }$  &   I   &   2.24    $\pm$   0.07    &   6.38    $\pm$   0.09    &   1   \\
HD 283572   &   o   &   G5  &   5770    &   0.38    &   5.78    $^{+    1.99    }_{ -1.31   }$  &   I   &   1.65    $\pm$   0.11    &   6.94    $\pm$   0.08    &   1   \\
RX J0423.7+1537 &   x   &   G5  &   5770    &   0.72    &   1   $^{+    0.36    }_{ -0.23   }$  &   V,I &   1.11:           &   7.48:           &   3   \\
IP Tau  &   c   &   M0  &   3850    &   0.24    &   0.34    $^{+    0.24    }_{ -0.12   }$  &   I   &   0.59    $\pm$   0.07    &   6.63    $\pm$   0.25    &   1   \\
DF Tau A   &   c   &   M1  &   3720    &   0.04    &   0.47    $^{+    0.30    }_{ -0.15   }$  &   I   &   0.48    $\pm$   0.05    &   6.28    $\pm$   0.17    &   1,8   \\
DF Tau B   &   c   &   M3.5  &   3420    &   0.04    &   0.53    $^{+    0.35    }_{ -0.18   }$  &   I   &   0.32    $\pm$   0.03    &   6.08    $\pm$   0.13    &   1,8   \\
NTTS 042417+1744    &   x   &   K1  &   5080    &   0.1 &   1.99    $^{+    0.82    }_{ -0.51   }$  &   I   &   1.51    $\pm$   0.14    &   6.92    $\pm$   0.13    &   1   \\
DI Tau AB   &   o   &   M0  &   3850    &   0.76    &   0.85    $^{+    0.9 }_{ -0.35   }$  &   I   &   0.56    $\pm$   0.06    &   6.11    $\pm$   0.21    &   1   \\
IQ Tau  &   c   &   M0.5    &   3785    &   1.25    &   0.53    $^{+    0.44    }_{ -0.2    }$  &   I   &   0.52    $\pm$   0.05    &   6.28    $\pm$   0.21    &   1   \\
UX Tau A    &   c   &   K5  &   4350    &   0   &   1.29    $^{+    0.5 }_{ -0.32   }$  &   I   &   1.12    $\pm$   0.13    &   6.43    $\pm$   0.17    &    1   \\
UX Tau B    &   o   &   M2  &   3580    &   0   &   0.15    $^{+    0.06    }_{ -0.04   }$  &   J   &   0.38    $\pm$   0.04    &   6.75    $\pm$   0.17    &  5    \\
%UX Tau C   &   o   &   M5  &   3240    &   0   &   0.06    $^{+    0.02    }_{ -0.01   }$  &   J   &   0.19    $\pm$   0.03    &   6.8 $\pm$   0.1 &    5   \\
FX Tau AB   &   c   &   M1  &   3720    &   2.24    &   0.89    $^{+    0.7 }_{ -0.32   }$  &   I   &   0.47    $\pm$   0.05    &   6.05    $\pm$   0.1 &   1   \\
DK Tau AB   &   c   &   K7  &   4060    &   0.35    &   0.9 $^{+    0.32    }_{ -0.21   }$  &   I   &   0.76    $\pm$   0.08    &   6.31    $\pm$   0.13    &   1   \\
V927 Tau AB &   c   &   M3  &   3470    &   1.4 &   0.6 $^{+    0.49    }_{ -0.22   }$  &   I   &   0.35    $\pm$   0.03    &   6.04    $\pm$   0.18    &   1   \\
%V927 Tau A &   c   &   M3  &   3470    &   1.4 &   0.31    $^{+    0.26    }_{ -0.12   }$  &   J   &   0.34    $\pm$   0.03    &   6.3 $\pm$   0.16    &       \\
%V927 Tau B &   c   &   M3.5    &   3420    &   1.4 &   0.21    $^{+    0.18    }_{ -0.08   }$  &   J   &   0.32    $\pm$   0.03    &   6.44    $\pm$   0.17    &       \\
NTTS 042835+1700    &   x   &   K5  &   4350    &   0.21    &   0.19    $^{+    0.19    }_{ -0.08   }$  &   I   &   0.82    $\pm$   0.09    &   7.57    $\pm$   0.24    &   1   \\
HK Tau AB   &   c   &   M0.5    &   3785    &   3.41    &   1   $^{+    1.78    }_{ -0.49   }$  &   I   &   0.52    $\pm$   0.05    &   5.97    $\pm$   0.27    &   1   \\
%HK Tau A   &   c   &   M1  &   3720    &   3.41    &   0.92    $^{+    1.63    }_{ -0.45   }$  &   J   &   0.47    $\pm$   0.04    &   5.91    $\pm$   0.37    &       \\
%HK Tau B   &   c   &   M2  &   3580    &   3.41    &   0.04    $^{+    0.08    }_{ -0.02   }$  &   J   &   0.36    $\pm$   0.05    &   7.48    $\pm$   0.42    &       \\
L 1551-51 AB    &   o   &   K7  &   4060    &   0   &   0.49    $^{+    0.14    }_{ -0.1    }$  &   I   &   0.79    $\pm$   0.08    &   6.71    $\pm$   0.16    &   1   \\
V827 Tau    &   o   &   K7  &   4060    &   0.28    &   0.87    $^{+    0.45    }_{ -0.26   }$  &   I   &   0.76    $\pm$   0.08    &   6.33    $\pm$   0.17    &   1   \\
V826 Tau AB &   o   &   K7  &   4060    &   0.08    &   0.5 $^{+    0.12    }_{ -0.09   }$  &   I   &   0.79    $\pm$   0.08    &   6.69    $\pm$   0.14    &   1   \\
V928 Tau AB &   o   &   M0.5    &   3785    &   1.87    &   0.82    $^{+    0.64    }_{ -0.29   }$  &   I   &   0.52    $\pm$   0.05    &   6.07    $\pm$   0.16    &   1   \\
RX J0432.8+1735 &   x   &   M2  &   3580    &   0.65    &   0.52    $^{+    0.78    }_{ -0.24   }$  &   V,J &   0.4 $\pm$   0.04    &   6.14    $\pm$   0.2 &   3    \\
%GG Tau AB   &   m   &   K7-M0   &   3950    &   0.76    &   0.79    $^{+    0.2 }_{ -0.15   }$  &   I   &   0.65    $\pm$   0.07    &   6.26    $\pm$   0.12    &   1   \\
GG Tau Aa   &   c   &   K7  &   4060    &   0.7 &   0.64    $^{+    0.16    }_{ -0.12   }$  &   I   &   0.78    $\pm$   0.08    &   6.52    $\pm$   0.13    &    6   \\
GG Tau Ab   &   c   &   M0.5    &   3720    &   3.2 &   0.57    $^{+    0.14    }_{ -0.1    }$  &   I   &   0.48    $\pm$   0.05    &   6.19    $\pm$   0.09    &   6    \\
%GG Tau Ba  &   c   &   M5  &   3240    &   0.6 &   0.07    $^{+    0.02    }_{ -0.01   }$  &   I   &   0.2 $\pm$   0.03    &   6.75    $\pm$   0.09    &    6   \\
%GG Tau Bb  &   c   &   M7  &   2860    &   0   &   0.009   $^{+    0.002   }_{ -0.002  }$  &   I   &   -   &   &  8 \\
GH Tau AB   &   c   &   M2  &   3580    &   1.04    &   0.98    $^{+    0.71    }_{ -0.34   }$  &   I   &   0.4 $\pm$   0.03    &   5.82    $\pm$   0.32    &   1   \\
%GH Tau A   &   c   &   M2  &   3580    &   1.04    &   0.44    $^{+    0.32    }_{ -0.15   }$  &   J   &   0.4 $\pm$   0.04    &   6.21    $\pm$   0.14    &       \\
%GH Tau B   &   c   &   M2  &   3580    &   1.04    &   0.39    $^{+    0.28    }_{ -0.13   }$  &   J   &   0.4 $\pm$   0.04    &   6.26    $\pm$   0.15    &       \\
DM Tau  &   c   &   M1  &   3720    &   0   &   0.16    $^{+    0.24    }_{ -0.07   }$  &   I   &   0.47    $\pm$   0.06    &   6.87    $\pm$   0.34    &   1   \\
CI Tau  &   c   &   K7  &   4060    &   1.77    &   0.96    $^{+    1.36    }_{ -0.44   }$  &   I   &   0.76    $\pm$   0.08    &   6.27    $\pm$   0.28    &   1   \\
NTTS 043124+1824    &   x   &   G8  &   5520    &   1.08    &   2.93    $^{+    3.21    }_{ -1.22   }$  &   I   &   1.55    $\pm$   0.22    &   7   $\pm$   0.16    &   1   \\
V1078 Tau   &   x   &   F8  &   6200    &   0.94    &   5.82    $^{+    2.91    }_{ -1.66   }$  &   V,J &   1.53    $\pm$   0.11    &   7.09    $\pm$   0.09    &    2   \\
DN Tau  &   c   &   M0  &   3850    &   0.49    &   0.79    $^{+    0.22    }_{ -0.16   }$  &   I   &   0.57    $\pm$   0.06    &   6.15    $\pm$   0.11    &   1   \\
LkCa 14 &   o   &   M0  &   3850    &   0   &   0.76    $^{+    0.22    }_{ -0.15   }$  &   I   &   0.57    $\pm$   0.06    &   6.16    $\pm$   0.11    &   1   \\
DO Tau  &   c   &   M0  &   3850    &   2.64    &   1.29    $^{+    1.15    }_{ -0.49   }$  &   I   &   0.56    $\pm$   0.05    &   5.93    $\pm$   0.15    &   1   \\
HV Tau AB   &   c   &   M1  &   3720    &   2.42    &   0.6 $^{+    0.22    }_{ -0.14   }$  &   I   &   0.48    $\pm$   0.05    &   6.17    $\pm$   0.11    &   1   \\
VY Tau AB   &   c   &   M0  &   3850    &   0.38    &   0.3 $^{+    0.31    }_{ -0.12   }$  &   I   &   0.59    $\pm$   0.07    &   6.7 $\pm$   0.3 &   1   \\
LkCa 15 &   c   &   K5  &   4350    &   0.62    &   0.85    $^{+    0.3 }_{ -0.2    }$  &   I   &   1.12    $\pm$   0.08    &   6.7 $\pm$   0.16    &   1   \\
IW Tau AB   &   o   &   K7  &   4060    &   0.83    &   0.99    $^{+    0.7 }_{ -0.34   }$  &   I   &   0.75    $\pm$   0.08    &   6.26    $\pm$   0.19    &   1   \\
LkH$\alpha$ 332 G2 AB   &   o   &   K7  &   4060    &   3.16    &   0.49    $^{+    0.3 }_{ -0.16   }$  &   V   &   0.79    $\pm$   0.08    &   6.7 $\pm$   0.24    &    1   \\
GO Tau  &   c   &   M0  &   3850    &   1.18    &   0.22    $^{+    0.18    }_{ -0.08   }$  &   I   &   0.59    $\pm$   0.07    &   6.91    $\pm$   0.28    &   1   \\
DR Tau  &   c   &   K7  &   4060    &   2:    &   1.97    $^{+    3.02    }_{ -0.92   }$  &   I   &   0.74    $\pm$   0.08    &   5.92    $\pm$   0.23    &   1   \\
UY Aur AB   &   c   &   K7  &   4060    &   2.05    &   1.41    $^{+    0.57    }_{ -0.35   }$  &   I   &   0.74    $\pm$   0.07    &   6.07    $\pm$   0.11    &   1   \\
RX J0452.5+1730 &   x   &   K4  &   4590    &   0:    &   0.39    $^{+    0.57    }_{ -0.18   }$  &   V   &   0.99    $\pm$   0.17    &   7.3 $\pm$   0.27    &   3    \\
GM Aur  &   c   &   K3  &   4730    &   0.14    &   1.23    $^{+    1.07    }_{ -0.47   }$  &   I   &   1.37    $\pm$   0.17    &   6.87    $\pm$   0.23    &   1   \\
LkCa 19 &   o   &   K0  &   5250    &   0   &   1.56    $^{+    0.5 }_{ -0.34   }$  &   I   &   1.28    $\pm$   0.1 &   7.17    $\pm$   0.1 &   1   \\
AB Aur  &   a   &   B9  &   10500   &   0.62    &   44.86   $^{+    20  }_{ -12 }$  &   I   &   2.58    $\pm$   0.14    &   6.68    $\pm$   0.08    &   1   \\
SU Aur  &   c   &   G2  &   5860    &   0.9 &   9.29    $^{+    2.26    }_{ -1.65   }$  &   I   &   1.88    $\pm$   0.1 &   6.8 $\pm$   0.08    &   1   \\
NTTS 045251+3016 AB &   x   &   K7  &   4060    &   0   &   0.54    $^{+    0.14    }_{ -0.1    }$  &   I   &   0.79    $\pm$   0.08    &   6.64    $\pm$   0.14    &   7    \\
RX J0457.0+1517 &   x   &   G0  &   6030    &   0.75    &   1.75    $^{+    0.6 }_{ -0.4    }$  &   V,J &   1.21    $\pm$   0   &   7.55    $\pm$   0.31    &    3   \\
RX J0457.5+2014 AB  &   x   &   K3  &   4730    &   0.05    &   0.86    $^{+    0.31    }_{ -0.2    }$  &   V,J &   1.22    $\pm$   0.11    &   7.09    $\pm$   0.13    &   3    \\
RW Aur A  &   c   &   K4  &   4590    &   0.32    &   1.72    $^{+    0.47    }_{ -0.33   }$  &   I   &   1.07    $\pm$   0.06    &   7.20    $\pm$   0.11    &   1,8   \\
\end{longtable}
}
\end{document}